\documentstyle[aps,multicol,amssymb,psfig]{revtex}

\begin{document}
\title{Size effects of pyroelectric coefficient and dielectric susceptibility in
ferroelectric thin films}
\author{M.D. Glinchuk and E.A. Eliseev}
\address{Institute for Materials Sciences, NASC of\\
Ukraine, Krjijanovskogo 3, 252180, Kiev, Ukraine}
\author{V. A. Stephanovich}
\address{Institute of Mathematics, University of Opole, Oleska 48,\\
45-052, Opole, Poland}
\date{\today}
\maketitle

\begin{abstract}
We calculate the pyrocoefficient, static dielectric susceptibility
profiles and its thickness dependence of ferroelectric thin films. 
Also, the temperature dependences of above quantities have been calculated.
For the calculations we use Landau phenomenological theory, leading to Lame 
equations. These equations subject to boundary conditions with different
extrapolation length on the surfaces have been solved numerically.
The divergency of pyroelectric
coefficient and static dielectric susceptibility in the vicinity of
thickness induced ferroelectric phase transition (i.e. at $l\approx l_c$ or $%
T\approx T_{cl}$) has been shown to be the most prominent size effect in
ferroelectric thin films.

We also calculate the dependence of critical 
thickness $l_c$ and critical temperature $T_{cl}$
on extrapolation length and film thickness. The
latter defines temperature - thickness phase diagram of the film with
paraelectric phase at $l\leq l_c$, $T\geq 0$ or $T\geq T_{cl}$ and
ferroelectric phase at $l>l_c$ and $T< T_{cl}$. 
%We show that in the vicinity of $l_c$ and $T_{cl}$ thickness and temperature dependence is 
%described
%approximately by $1/\sqrt{l-l_c}$, $1/\sqrt{T_{cl}-T}$ for pyrocoefficient
%and $1/\left| l_c-l\right| $, $1/\left| T_{cl}-T\right| $ for dielectric
%susceptibility, the accuracy of this approximation in temperature dependence
%being large in broad ehough temperature region, while in thickness
%dependence this approximation can be suitable in the vicinity of the
%critical thickness $l_c$ only. 
We compare our theory with available experimental data.

\end{abstract}

\pacs{PACS numbers: 6865.+g, 7755.+f, 7784.-s}

\begin{multicols}{2}

\narrowtext

\section{Introduction}

Ferroelectric thin films attract much attention of the scientists due to
important fundamental problems related to size effects resulting in
drastic difference between physical properties of the films and of the
bulk ferroelectrics (see e.g. \cite{1}). On the other hand in the recent
years a lot of efforts have been spared to the application of ferroelectric
thin films in pyroelectric and piezoelectric sensors, in some types of random
access memory devices, optical memories, microswitches etc. \cite{2,3}. The
prospect of these and other applications depends strongly on the solution of
fundamental scientific problems of thin films anomalous properties and their
dependence on the film thickness. The latter seems to be especially
important since the film thickness can be easily adjusted, this adjustment can lead
to essential change of the physical properties. In particular, to obtain thin
film pyroelectric sensors or capacitors with high performance one has to
know the film thickness that gives maximal value of pyrocoefficient or
dielectric permittivity respectively. Keeping in mind that recently giant
dielectric response (several hundred thousands) was observed in
ferroelectric thin film multilayers \cite{4} and its origin was shown to be
thickness induced ferroelectric phase transition \cite{5}, one can suppose
the existence of anomalous dielectric response and other properties
peculiarities in the single film with some specially adjusted thickness. 
The critical thickness corresponding to disappearance of spontaneous polarization
(thickness induced ferroelectric phase transition) had been calculated recently
in several papers (see e.g. \cite{6,7}) in the framework of phenomenological
Landau theory. There are only few works devoted to dielectric permittivity
calculations (see e.g. \cite{8}). To the best of our knowledge there are no
calculations of pyroelectric coefficient profile and its thickness
dependence in ferroelectric thin films. Besides that, all the calculations were
performed in supposition that extrapolation lengths (i.e. boundary
conditions) on both film surfaces are the same. This seems to be
rough approximation due to different conditions on the film surfaces
(e.g. substrate and air). Moreover, the greatest part of calculations of
ferroelectric thin films properties were performed numerically mainly for
the parameters of BaTiO$_3$ or PbTiO$_3$ films (see e.g. \cite{9,10}), which
makes it difficult to apply the results for arbitrary ferroelectric film
and to extract general features of size effects in the films.

In the present paper we perform the analytical calculations of pyroelectric
coefficient and dielectric permittivity in ferroelectric thin film in the
framework of the phenomenological theory. The profiles, thickness and
temperature dependences of these quantities were obtained by analytical
solution of Lame-type equations with different extrapolation lengths (i.e.
different boundary conditions) on the surfaces. The theory predicts the divergency
of dielectric permittivity and pyrocoefficient in the film with critical
thickness. We compare our theory with available experimental data.

\section{Theory. General equations}

In what follows we shall consider a film polarized along $z$ direction
normal to the surfaces of the film, i.e. polarization $P\equiv P_z$, $%
P_x=P_y=0$. This type of polarization can appear as a result of
self-polarization of a film grown under special technological conditions
without application of an external electric field \cite{11}.

In the phenomenological theory $P_z(z)\equiv P$ can be obtained by the
minimization of free energy functional
\end{multicols}
\widetext
\noindent\rule{20.5pc}{0.1mm}\rule{0.1mm}{1.5mm}\hfill

\begin{equation}
F=\int\limits_0^l\left( \frac{\alpha P^2}2+\frac{\beta P^4}4+\frac \gamma 2%
\left( \frac{dP}{dz}\right) ^2-PE\right) dz+\frac \gamma 2\left[ \left. 
\frac{P^2}{\delta _1}\right| _{z=0}+\left. \frac{P^2}{\delta _2}\right|
_{z=l}\right]  \label{1}
\end{equation}
\hfill\rule[-1.5mm]{0.1mm}{1.5mm}\rule{20.5pc}{0.1mm} 
\begin{multicols}{2}
\narrowtext
where $E\equiv E_z$ is an external electric field, $l$ is the film thickness, $%
\delta _1$ and $\delta _2$ are extrapolation lengths on the film surfaces.
The coefficients $\alpha $, $\beta $, $\gamma $ are those of bulk material
renormalized by internal mechanical stresses originated from mismatch of
substrate and film lattice constants, expansion coefficients, growth
imperfections (see \cite{12,5}) and depolarizing field. Since this
renormalization can result into transformation of first order phase
transition into that of a second order \cite{12}, Eq.(\ref{1}) can be valid for the
materials both with first and second order bulk phase transition.

\begin{figure}[th]
\vspace*{3mm}
\centerline{\centerline{\psfig{figure=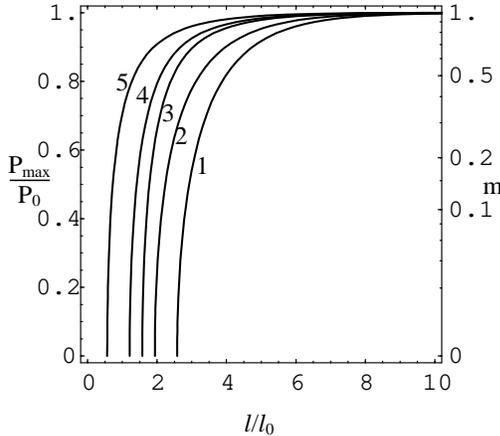,width=0.8\columnwidth}}}
\caption{The thickness dependence of $P_{max}$ and $m$
for several values of extrapolation length. Curves 1, 2, 3, 4 and 5
correspond to extrapolation lenghts $(\protect\delta _{01},\,\protect\delta
_{02})=(0.1,\,0.5),$ (0.1, 2); (0.1, 10), (0.5, 10) and (2, 10)
respectively. Arrows correspond to critical thickness.}
\end{figure}

The minimization of the functional (\ref{1}) 
implies following Euler-Lagrange equation
\begin{mathletters}
\begin{equation}
\frac{\partial F}{\partial P}-\frac d{dz}\frac{dF}{dP^{^{\prime }}}=0,
P^{^{\prime }}\equiv \frac{dP}{dz}  \label{2a} 
\end{equation}
and boundary conditions

\begin{equation}
\left. \frac{dP}{dz}\right| _{z=0}=\left. \frac P{\delta _1}\right|
_{z=0}; \left. \frac{dP}{dz}\right| _{z=l}=-\left. \frac P{\delta
_2}\right| _{z=l}  \label{2b}  
\end{equation}
\end{mathletters}
It is seen from Eqs.(\ref{2b}) that extrapolation lengths determine the crossing
points $z_1$ and $z_2$ between $z$ axis and tangent lines to $P(z)$ curve at
the points $z=0$ and $z=l$ respectively, i.e. $\delta _1=-z_1$ and $\delta
_2=z_2-l$ . From this geometric definition of $\delta _1$ and $\delta _2$
one can see that $\delta _1>0$, $\delta _2>0$ when polarization on the
surfaces is smaller than that in the film and both $\delta _1$ and $\delta
_2 $ are negative in opposite case. Since the conditions on two surfaces can
be strongly different (e.g. substrates, electrodes etc.) it may be possible
to have different signs of $\delta _1$ and $\delta _2$, e.g. $\delta _1>0$
if $P(z=0)<P(z)$ and $\delta _2<0$ if $P(z=l)>P(z)$. In what follows we
shall consider the case $\delta _1>0$, $\delta _2>0$. Eqs. (\ref{1}) and (\ref{2a}) lead
to the following equation for equilibrium inhomogeneous
polarization in the film:

\begin{equation}
\alpha P+\beta P^3-\gamma \frac{d^2P}{dz^2}-E=0  \label{3} 
\end{equation}
which has to be solved subject to the boundary conditions (\ref{2b}).

Since $\alpha =\alpha _0(T-T_c)$, where $T_c$ is the temperature of
ferroelectric phase transition of the thick film (see e.g. \cite{5}),
inhomogeneous polarization of the film $P(z)$ has to depend on $T$ and $E$.
This lead to inhomogeneous pyrocoefficient $\Pi (z)=(dP(z)/dT)_{P=P_S}$ and
linear dielectric susceptibility $\chi (z)=(dP(z)/dE)_{P=P_S}$ where $P_s$
is spontaneous polarization defined by Eq.(\ref{3}) at $E=0$ and boundary
conditions (\ref{2b}). The differentiation of Eqs. (\ref{3}) and (\ref{2b}) gives
following differential equations for pyrocoefficient and dielectric
susceptibility

\begin{mathletters}
\begin{equation}
(\alpha +3\beta P_s^2)\Pi -\gamma \frac{d^2\Pi }{dz^2}+\alpha _0P_s=0; 
\label{4}  
\end{equation}
\begin{equation}
\left. \frac{d\Pi }{dz}\right| _{z=0}=\left. \frac \Pi {\delta _1}\right|
_{z=0};\left. \frac{d\Pi }{dz}\right| _{z=l}=-\left. \frac \Pi {\delta
_2}\right| _{z=l}  \label{4b} 
\end{equation}
\end{mathletters}

\begin{mathletters}
\begin{equation}
(\alpha +3\beta P_s^2)\chi -\gamma \frac{d^2\chi }{dz^2}-1=0;  \label{5a}
\end{equation}
\begin{equation}
\left. \frac{d\chi }{dz}\right| _{z=0}=\left. \frac \chi {\delta _1}\right|
_{z=0};\left. \frac{d\chi }{dz}\right| _{z=l}=-\left. \frac \chi {\delta
_2}\right| _{z=l}  \label{5}  
\end{equation}
\end{mathletters}
The Eqs. (\ref{4}) and (\ref{5}) define, respectively, pyroelectric coefficient and
dielectric susceptibility profiles, i.e. $\Pi (z)$ and $\chi (z)$, as well
as their dependences on temperature and thickness of the film. Mean values
of pyroelectric coefficient $\bar \Pi $ and dielectric susceptibility $%
\overline{\chi }$ can be calculated as follows

\begin{equation}
\overline{\Pi }=\frac 1l\int\limits_0^l\Pi (z)dz,~\overline{\chi }=\frac
1l\int\limits_0^l\chi (z)dz.  \label{6}  
\end{equation}

\begin{figure}[th]
\vspace*{-6mm}
\centerline{\centerline{\psfig{figure=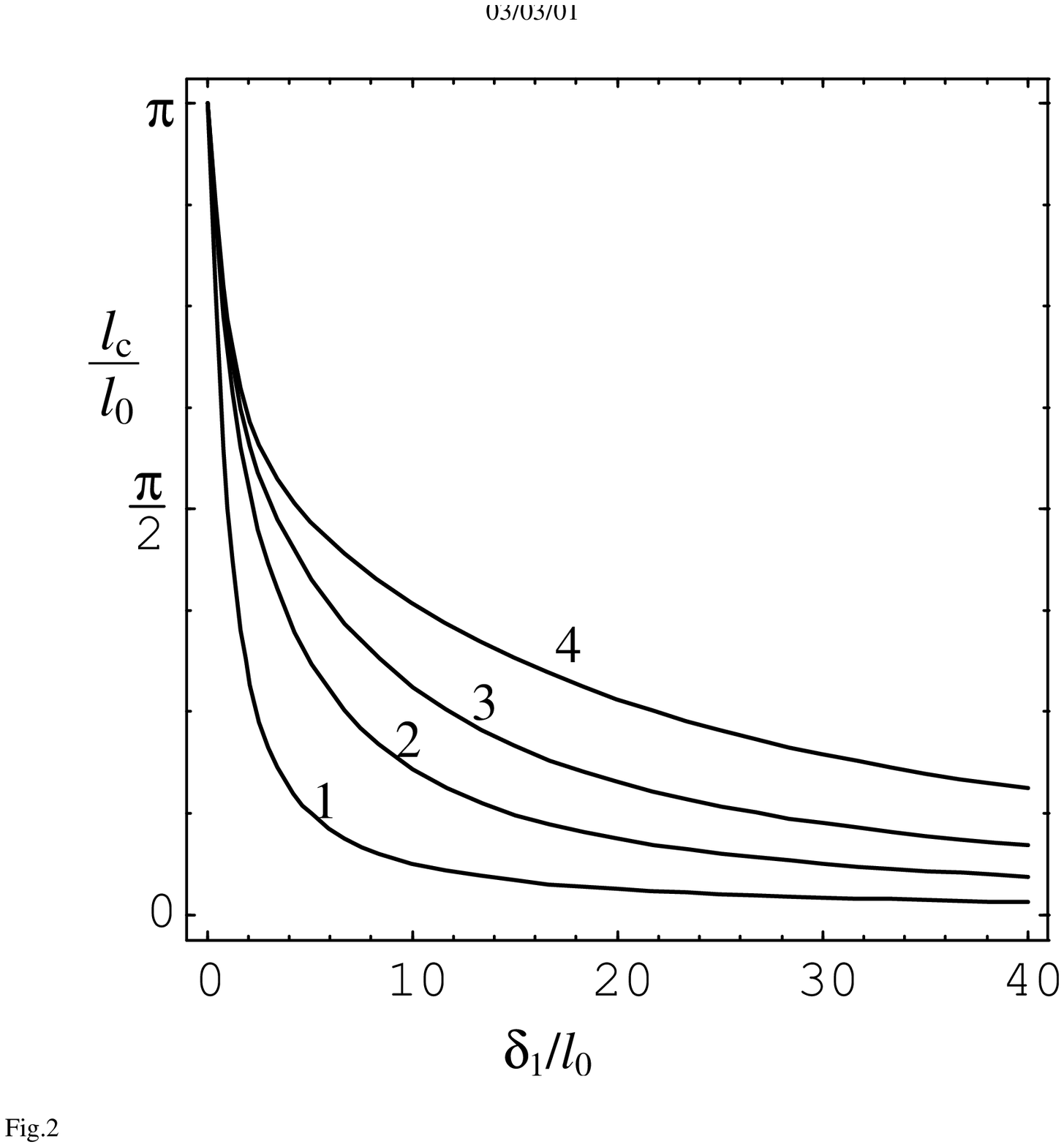,width=0.8\columnwidth}}}
%\vspace*{mm}
\caption{The dependence of the critical thickness on the extrapolation
length. Curves 1, 2, 3 and 4 correspond to $\protect\delta _{02}=\protect%
\delta _{01}$, $\protect\delta _{02}=0.2\,\protect\delta _{01}$, $\protect%
\delta _{02}=0.1\,\protect\delta _{01}$ and $\protect\delta _{02}=0.05\,%
\protect\delta _{01}$ respectively.}
\end{figure}

\section{Inhomogeneous spontaneous polarization}

Spontaneous polarization can be found by solution of Eq.(\ref{3}) at $E=0$ with
boundary conditions (\ref{2b}). 
In dimensionless variables
\begin{mathletters}
\begin{equation}
P_1=\frac{P_z}{P_0}, z_1=\frac z{l_0}, l_1=\frac l{l_0},
\delta _{01,2}=\frac{\delta _{1,2}}{l_0},  \label{7} 
\end{equation}
where $l_0=\sqrt{-\gamma /\alpha },$ $P_0=\sqrt{-\alpha /\beta }$ ($\alpha
=\alpha _0(T-T_c)$, $T\leq T_c$) is the homogeneous polarization of a thick
film (where the contribution of gradient can be neglected), the equation for
spontaneous polarization of thin film has the form:

\begin{equation}
\frac{d^2P_1}{dz_1^2}-P_1-P_1^3=0  \label{7b}  
\end{equation}
\end{mathletters}
The solution of this nonlinear equation subject to above boundary conditions 
%with restricted $P_{1z}$ and $%
%dP_{1z}/dz$ values (which are realized when polarization in the thick film
%is the largest one, i.e. $P_0>P_z$ \cite{13}) 
can be written in terms of elliptic sine (sn(...)) (see e.g. \cite{14,15})

\begin{equation}
P_{1z}=\sqrt{\frac{2m}{1+m}}{\rm sn}\left( \frac{z_1+z_0}{\sqrt{1+m}},m\right) 
\label{8}  
\end{equation}
where $m$ and $z_0$ are the parameters defined by the boundary conditions
and film thickness.

In particular, value of $z_0$ can be obtained from 
Eqs. (\ref{8}) and (\ref{2b}) at $z_1=0$. It can be written as 
\begin{equation}
z_0=\sqrt{1+m}~F(\arcsin (f_1),m)  \label{9} 
\end{equation}
Values of parameter $m$ can be found from the equation
\end{multicols}
\widetext
\noindent\rule{20.5pc}{0.1mm}\rule{0.1mm}{1.5mm}\hfill
 
\begin{equation}
l_1=2\sqrt{1+m}\left( 2K(m)-F(\arcsin (f_1),m)-F(\arcsin (f_2),m)\right)
\label{10}
\end{equation}

Here

\begin{equation}
f_i\equiv f(m,\delta _{0i})=\sqrt{\frac{1+m}{2m}\left( 1+\frac 1{\delta
_{0i}^2}-\sqrt{\left( 1+\frac 1{\delta _{0i}^2}\right) ^2-\frac{4m}{\left(
1+m\right) ^2}}\right) }  \label{11}
\end{equation}
\hfill\rule[-1.5mm]{0.1mm}{1.5mm}\rule{20.5pc}{0.1mm} 
\begin{multicols}{2}
\narrowtext
Note, that to obtain Eq.(\ref{10}) we used Eqs. (\ref{8}) and (\ref{2b}) at $z=l$. 
In Eqs.(\ref{9},\ref{10}) 
$K(m)$ and $F(\varphi ,m)$ are complete and incomplete elliptic
integrals of the first kind respectively \cite{14,15}. The Eqs. (\ref{8})-(\ref{11}) are
general formulas for calculation of inhomogeneous polarization of ferroelectric
thin films.

It should be noted, that nontrivial solution of equation (\ref{3}) in the temperature
region $T>T_c$ and zero external field $E=0$ can not satisfy boundary
conditions (\ref{2a}) with positive extrapolation lengths, so only trivial
solution $P_s=0$ is possible in this case.

\section{Thickness induced ferroelectric phase transitions}

The analysis of Eq.(\ref{10}) has shown that values of $m$ lie in the interval $%
\left[ 0,1\right] $, $m=1$ being the value for thick film. We report 
maximal polarization and $m$ values (respectively on left and right
side scales) on Fig.1 as the functions of a film thickness. One can see that at $%
l/l_0\approx 8 P_m\approx P_0$, i.e. maximal polarization in the film
is close to that of a thick film. It follows from Eq.(\ref{8}), that at $m=0$
polarization equals zero. This situation appears at some critical thickness $%
l_c$, that was found from Eqs. (\ref{10}), (\ref{11}) at $m=0$. For the considered case $%
\delta _{01}\geq 0$, $\delta _{02}\geq 0$. It can be written in the form

\end{multicols}
\widetext
\noindent\rule{20.5pc}{0.1mm}\rule{0.1mm}{1.5mm}\hfill
\begin{equation}
l_{1c}=\frac{l_c}{l_0}=\left( \pi -\arcsin \left( \frac{\delta _{01}}{\sqrt{%
1+\delta _{01}^2}}\right) -\arcsin \left( \frac{\delta _{02}}{\sqrt{1+\delta
_{02}^2}}\right) \right)  \label{12}
\end{equation}
\hfill\rule[-1.5mm]{0.1mm}{1.5mm}\rule{20.5pc}{0.1mm} 
\begin{multicols}{2}
\narrowtext

Eq.(\ref{12}) represents the dimensionless critical thickness at which thickness
induced ferroelectric phase transition occurs, i.e. ferroelectric phase
disappears in the films thinner than critical thickness $l_c$.

One can see from Eq.(\ref{12}) that maximal value of $l_{1c}$ ($l_{1c}=\pi $)
corresponds to $\delta _{01}=\delta _{02}=0$, i.e. for $\left. P(z)\right|
_{z=0}=\left. P(z)\right| _{z=l}=0$ (see Eq. (\ref{2b})). Increase of $\delta
_{01} $ value leads to decrease $l_{1c}$ so that at $\delta
_{01,2}\to \infty  l_{1c}\to 0$. In the limit $\delta
_{01}\to \infty $, $\delta _{02}\to 0$ (polarization on one
of the surfaces equals zero and acquires some maximal value on the other surface) 
$l_{1c}\to \pi /2$ (see Eq.(\ref{12})). In Fig.2 the values of $l_{1c}$ for
intermediate $\delta _{01}\neq \delta _{02}$ are reported.
In particular it follows from this Figure that $\delta _{01}=40,$ $\delta
_{02}=2$ is far enough from the limit $l_{1c}\to \pi /2$, which can
be achieved at smaller $\delta _{02}$ values.

The thickness induced ferroelectric phase transition can occur also in
the film with some arbitrary thickness $l\neq l_c$ at critical temperature $%
T=T_{cl}$. Qualitatively this follows from the fact that $m$ depends
both on $l$ and $T$ so that $m=0$ can be obtained at some special (critical)
values of film thickness or temperature. Quantitatively $T_{cl}$ as a
function of film thickness and extrapolation length can be obtained
from Eq.(\ref{12}) by substituting $l_0=\sqrt{\gamma /\left( \alpha _0T_c(1-\tau
_{cl})\right) }$, $\tau _{cl}=T_{cl}/T_c$ and $l_c=l$. This transforms
Eq.(\ref{12}) into that for $\tau _{cl}$, namely
\end{multicols}
\widetext
\noindent\rule{20.5pc}{0.1mm}\rule{0.1mm}{1.5mm}\hfill
\begin{mathletters}
\begin{equation}
\lambda \sqrt{1-\tau _{cl}}=\pi -\arcsin \frac{d_1\sqrt{1-\tau _{cl}}}{\sqrt{%
1+d_1^2(1-\tau _{cl})}}-\arcsin \frac{d_2\sqrt{1-\tau _{cl}}}{\sqrt{%
1+d_2^2(1-\tau _{cl})}},  \label{13a}
\end{equation}
where

\begin{equation}
\lambda =l\sqrt{\frac{\alpha _0T_c}\gamma }=\frac l{l_0(T=0)},
d_{1,2}=\delta _{1,2}\sqrt{\frac{\alpha _0T_c}\gamma }=\frac{\delta _{1,2}}{%
l_0(T=0)}  \label{13b}
\end{equation}
\hfill\rule[-1.5mm]{0.1mm}{1.5mm}\rule{20.5pc}{0.1mm} 
\begin{multicols}{2}
\narrowtext

are renormalized dimensionless thickness and extrapolation lengths (compare
(\ref{7}) and (\ref{13b})). Expression (\ref{13a}) determines thickness dependence of
dimensionless temperature $\tau _{cl}$ of thickness induced ferroelectric
phase transition. We show it in Fig.3 for several values of $d_1$ and $%
d_2$. It is seen that the limit of thick film ($T_{cl}\simeq T_c)$
for all the curves corresponds to $l\gtrsim 15l_0(T=0)$. However this limit
can be achieved for thinner film ( $l\gtrsim 2l_0(T=0)$) in the case of
large enough $d_1$ and $d_2$ (see curve 1). The large $d_{1,2}$
can be related to large $\delta _{1,2}$ and/or $T_c$ values
and/or small $\gamma $ value (small contribution of polarization gradient),
which can be the consequence of strong enough ferroelectric phase transition
in a thick film. Note that values of critical thickness shown in Fig.1, 2
are given in the units $l_0=\sqrt{\gamma /\left( \alpha _0(T_c-T)\right) }$,
while in the Fig.3 they are given in the units $l_0(T=0)$.

Therefore ferroelectric phase transition in a film can be reached by
variation of the film thickness at some fixed temperature or by changing the
temperature of the film with fixed thickness. As a matter of fact the curves
in Fig.3 determine phase boundaries between paraelectric and ferroelectric
phases in coordinates temperature - thickness of the film. Nanely, at $l\leq
l_c$ paraelectric phase (polarization $P=0$) exists for all temperatures,
while at $l>l_c$ it can be both ferroelectric phase ($P\neq 0$) at $%
T<T_{cl}$ and paraelectric phase at $T\geq T_{cl}$.

\section{Pyroelectric coefficient and dielectric susceptibility profiles and
thickness dependence}

\subsection{Pyroelectric coefficient}

The solution of Eq. (\ref{4}) with the boundary conditions (\ref{4b}) defines the
distribution of pyroelectric coefficient $\Pi $ and has the following form
(see Appendix 1 for details):
\begin{equation}
\Pi (x)=C_1^{py}y_1(x)+C_2^{py}y_2(x)+y_3^{py}(x)  \label{14}  
\end{equation}
where function $y_i(x)$ have the form: 
\end{multicols}
\widetext
\noindent\rule{20.5pc}{0.1mm}\rule{0.1mm}{1.5mm}\hfill
\end{mathletters}
\begin{mathletters}
\begin{equation}
y_1(x)={\rm cn}(x,m){\rm dn}(x,m),  \label{15}
\end{equation}

\begin{equation}
y_2(x)=\left( x-\frac{1+m}{1-m}E(am(x),m)\right) \frac{y_1(x)}{1-m}+y_0(x)%
\frac{1+m^2-m(1+m)\,y_0^2(x)}{\left( 1-m\right) ^2},  \label{15b}
\end{equation}

\begin{eqnarray}
y_3^{py}(x) &=&-\Pi _m\left( \left( x-\frac{2\,E(am(x),m)}{1-m}\right)
y_1(x)+y_0(x)\frac{1+m-2m\,\,y_0^2(x)}{1-m}\right)  \nonumber \\
\Pi _m &=&\Pi _0\frac{\sqrt{2m\left( 1+m\right) }}{1-m}, \Pi _0=\frac{dP_0}{%
dT}  \label{15c}
\end{eqnarray}
\hfill\rule[-1.5mm]{0.1mm}{1.5mm}\rule{20.5pc}{0.1mm} 
\begin{multicols}{2}
\narrowtext

Here $\Pi _0$ is a thick film pyroelectric coefficient and the folowing
notations are introduced: 
\[
x=\frac{z_1+z_0}{\sqrt{1+m}}, y_0(x)={\rm sn}(x,m). 
\]
In Eqs.(\ref{15b}, \ref{15c}) $E(\varphi ,m)$ and ${\rm am}(x)$ are incomplete elliptic integral
of the second kind and elliptic amplitude function respectively \cite{14,15}.
We give the explicit expressions for $C_1^{py}$ and $C_2^{py}$ in Appendix 1 due to 
to their cumbersome form. These expressions along with Eqs.
(\ref{14}-\ref{15}) define the profile and temperature dependence of pyroelectric
coefficient of a thin film. 

\begin{figure}[th]
\vspace*{-1mm}
\centerline{\centerline{\psfig{figure=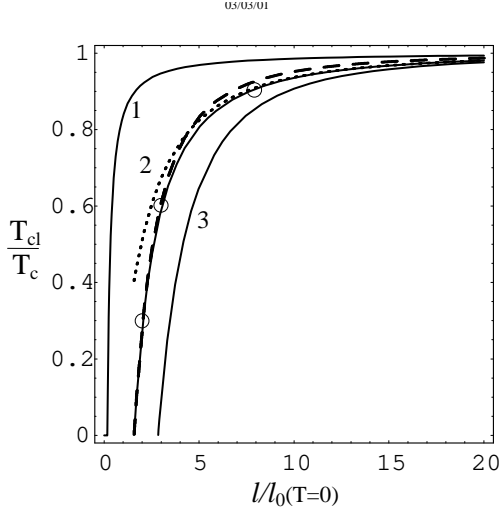,width=0.8\columnwidth}}}
\vspace*{1mm}
\caption{The thickness dependence of the critical temperature of thickness
induced ferroelectric phase transition for different dimentionless
extrapolation lengths: $(d_1,\,d_2)=(10,\,15)$, (2, 0.5), (0.2, 0,1)
respectively for solid lines 1, 2 and 3. The dashed and dotted lines
illustrate the approximate dependencies given by Eqs.{(\ref{30})} and {(\ref
{30a})} respectively with the same parameters as for the curve 2.}
\end{figure}

This coefficient profiles are reported in Fig.4
for several $m$ values. One can see from Fig.1 that $m$ can be
considered as a characteristic of a film thickness: $m\approx 1$
corresponds to the thickest film, while $m\to 0$ corresponds to
the film with critical thickness, i.e. to the thinnest possible film. One can see
from Fig.4 that for thick enough films pyroelectric coefficient profile has
two maxima in the vicinity of a film surfaces and a flat minimum in the
middle part of the film. The value of $\Pi (z)$ in this minimum depends on a
film thickness: it is close to pyroelectric coefficient of a thick film for
the thickest film (see curves 6 and 7) and it increases with a film
thickness decrease (see curve 5) up to complete disappearance of the minimum
and its transformation into a flat maximum for some intermediate film
thickness (see curves 3, 4). For thinner films pyroelectric coefficient
profile has one maximum. Its height and sharpness increases with a film
thickness decrease (see curves 1, 2). The strong increase of maximal $\Pi
(z) $ value for small $m$ speaks in favour of possible anomaly
of pyroelectric coefficient in the film with critical thickness. To check
this supposition we performed the calculations of the mean pyroelectric
coefficient $\overline{\Pi }$ on the base of Eq.(\ref{6}). 
With respect to Eq.(\ref{14}) one
can write $\overline{\Pi }$ in the following form (see Appendix 2 for
details):

\begin{equation}
\bar{\Pi}=C_1^{py}\bar{y}_1+C_2^{py}\bar{y}_2+\bar{y}_3^{py}  \label{16}
\end{equation}
where mean values of functions $y_i(x)$ have the form: 

\end{multicols}
\widetext
\noindent\rule{20.5pc}{0.1mm}\rule{0.1mm}{1.5mm}\hfill
\end{mathletters}
\begin{mathletters}
\begin{eqnarray}
\overline{y}_1 &=&\frac{f_2-f_1}{x_2-x_1},  \label{17a} \\
\overline{y}_2 &=&\frac{f_2\left( x_2-\frac{1+m}{1-m}\left( e_2-\frac{\sqrt{%
1+m}}{\delta _{02}}\right) \right) -f_1\left( x_1-\frac{1+m}{1-m}\left( e_1+%
\frac{\sqrt{1+m}}{\delta _{01}}\right) \right) }{\left( x_2-x_1\right)
\left( 1-m\right) },  \label{17b} \\
\overline{y}_3^{py} &=&\Pi _m\frac{f_2\left( 2\left( e_2-\frac{\sqrt{1+m}}{%
\delta _{02}}\right) -\left( 1-m\right) x_2\right) -f_1\left( 2\left( e_1+%
\frac{\sqrt{1+m}}{\delta _{01}}\right) -\left( 1-m\right) x_1\right) }{%
\left( x_2-x_1\right) \left( 1-m\right) }  \label{17c}
\end{eqnarray}
\hfill\rule[-1.5mm]{0.1mm}{1.5mm}\rule{20.5pc}{0.1mm}
\begin{multicols}{2}
\narrowtext

Here the following notations are introduced: 
\begin{eqnarray*}
x_1 &=&F(\arcsin (f_1),m), x_2=2K(m)-F(\arcsin (f_2),m) \\
e_1 &=&E(\arcsin (f_1),m), e_2=2E(m)-E(\arcsin (f_2),m)
\end{eqnarray*}
and $E(\varphi ,m)$ is the complete elliptic integral of the second kind 
\cite{14,15}. For the sake of illustration we depicted thickness dependence
of inverse mean pyrocoefficient $\overline{\Pi }$ in Fig.5 for several $%
\delta _{01},\ \delta _{02}$ values. One can see that at $l=l_c$ $\Pi _0%
\overline{/\Pi }\to 0$, i.e. there is the divergency of pyroelectric
coefficient for a film with critical thickness.

Temperature dependence of pyroelectric coefficient in the film with
arbitrary thickness can be derived on the base of obtained expressions
taking into account that the relation between the film thickness and
parameter $m$ (see Eq.(10)) depends on $T$ via parameter $l_0(T)=\sqrt{%
\gamma /\left( \alpha _0(T_c-T)\right) }\equiv l_0(\tau =0)/\sqrt{1-\tau }$, 
$\tau =T/T_c$ in variables given by (\ref{13b}). 
In these variables Eq.(\ref{10}) can be rewritten as
\end{multicols}
\widetext
\noindent\rule{20.5pc}{0.1mm}\rule{0.1mm}{1.5mm}\hfill

\end{mathletters}
\begin{eqnarray}
\lambda \sqrt{1-\tau } &=&\sqrt{1-m}\left[ 2K(m)-F\left( \arcsin \left(
f\left( m,d_1\sqrt{1-\tau }\right) \right) ,m\right) \right. -  \nonumber \\
&&\left. -F\left( \arcsin \left( f\left( m,d_2\sqrt{1-\tau }\right) \right)
,m\right) \right]  \label{18}
\end{eqnarray}
\hfill\rule[-1.5mm]{0.1mm}{1.5mm}\rule{20.5pc}{0.1mm}
\begin{multicols}{2}
\narrowtext

This expression defines the temperature dependence of parameter $m$. This
makes possible to calculate the temperature dependence of pyroelectric
coefficient on the base of Eq.(\ref{16}). The results are reported in Fig.6 by
solid lines for several film thickness. The divergency of $%
\overline{\Pi }(T)$ at $T=T_{cl}$ can be seen, the $T_{cl}$ values for the considered
film thickness are marked by open circles in Fig.3.

\subsection{Dielectric susceptibility}

Contrary to pyroelectric coefficient that exist only in ferroelectric phase,
dielectric response is the property of both ferroelectric and paraelectric
phase.

Let us begin with its calculation in ferroelectric phase that exist at $\dot
l>l_c$ and $T<T_{cl}$. In such a case it can be calculated similarly to
pyroelectric coefficient on the base of Eq.(\ref{5a},\ref{5}) (see Appendix 1 for
details): 
\begin{equation}
\chi (x)=C_1^{ch}y_1(x)+C_2^{ch}y_2(x)+y_3^{ch}(x)  \label{19}
\end{equation}
where 
\begin{equation}
y_3^{ch}(x)=-2\chi _0(1+m)\frac{1+m-2m\,y_0(x)}{(1-m)^2},  \label{20}
\end{equation}
$\chi _0=-1/2\alpha $ is thick film susceptibility. Coefficients $C_1^{ch}$
and $C_2^{ch}$ differ from $C_1^{py}$ and $C_2^{py}$ in Eq.(\ref{14}) because of
the difference between $y_3^{py}(x)$ and $y_3^{ch}(x)$ (see Appendix 1).
Therefore profiles of dielectric susceptibility can be described by Eq.(\ref{19})
with respect to Eqs.(\ref{15}, \ref{20}) and the values of $C_{1,2}^{ch}$ given in
Appendix 1.

The dielectric susceptibility profiles calculated on the base of the
equation (\ref{19}) are reported in Fig.7 for several values of parameter $m$.
Comparison of Fig.7 with Fig.4 shows that the profiles of
pyrocoefficient and dielectric susceptibility looks qualitatively similar with
the same peculiarities of $\chi (z)$ for thicker and thinner films as
it was described in previous subsection. However quantitatively the
behaviour of $\chi (z)$ for the thinner films differs from that for $\Pi
(z)$: the rate of $\chi (z)$ increase with film thickness decrease is
several times larger then the rate of $\Pi (z)$ increase (compare the Figs.
7 and 4). This speaks in favour of statement that dielectric susceptibility
is more sensitive to size effects than pyroelectric coefficient. To show
that the increase of $\chi $ in the thinner film is related to approaching
to thickness induced ferroelectric phase transition we performed the
calculation of mean dielectric susceptibility $\overline{\chi }$ similarly
to $\bar \Pi $ calculations (see Eq.(\ref{16}))
\begin{mathletters}
\begin{equation}
\overline{\chi }=C_1^{ch}\bar{y}_1+C_2^{ch}\bar{y}_2+\bar{y}_3^{ch}. 
\label{21a}  
\end{equation}
Integration of the third term in Eq.(\ref{19}) yields following expression
(see Appendix 2): 
\begin{equation}
\overline{y}_3^{ch}=\frac{\chi _0}{x_2-x_1}\frac{1+m}{1-m}\left( x_2-x_1-%
\frac{2\left( e_2-e_1\right) }{1-m}\right) .  \label{21b}
\end{equation}
\end{mathletters}
This expression along with Eqs.(\ref{17a}, \ref{17b}) defines the mean value of dielectric
susceptibility $\chi (z)$. It is shown on Fig.8 by right branches ($l>l_c$) 
of solid lines for several values of $\delta _{01}$ and $\delta _{02}$.
It is seen that at critical thickness marked by arrows $\chi _0/%
\overline{\chi }=0$ i.e. $\overline{\chi }\to \infty $ at $%
l\to l_c$ so that we have dielectric susceptibility
divergency in the film with critical thickness. Temperature dependence of
susceptibility can be calculated with the help of the same formulas (Eq.(\ref{18}))
keeping in mind that parameter $m$ depends also on temperature. 
The results of calculations are shown as the left branches ($T<T_{cl}$) of
solid lines in Fig.9. The divergency of $\chi (\tau )$ at $\tau =\tau
_{cl}=T_{cl}/T_c$ is seen clearly, the values of $\tau _{cl}$ in the
considered films being marked by open circles in Fig.3. Comparison of
pyroelectric coefficient and dielectric susceptibility temperature
dependencies (compare Fig.6 and Fig.9) shows that the rate of
approaching to infinity as $T\to T_{cl}$ is smaller
for pyroelectric coefficient than for susceptibility and $\chi _0/\overline{%
\chi }$ is linear function of $T/T_c$ contrary to $\Pi _0/\overline{\Pi }$.

Let us proceed now to calculation of susceptibility in paraelectric phase,
i.e. at $l<l_c$ and $T=const$ or $T>T_{cl}$ ($T_{cl}\leq T_c$ see Fig.3) and 
$l=const$. In such a case $P_S=0$ so that Eq.(\ref{5a}) gives following
equation for susceptibility

\begin{equation}
\alpha \chi -\gamma \frac{d^2\chi }{dz^2}=1,  \label{22}
\end{equation}
which has to be solved subject to boundary conditions (\ref{5}).

The equation (\ref{22}) is ordinary differential equation of the second order
with constant coefficients and its solution can be expressed via trigonometric
(when $\gamma /\alpha <0$) or hyperbolic functions (when $\gamma /\alpha >0$),
while in ferroelectric phase
the coefficient of Eq.(\ref{5a}), including $P_S^2,$ dependends on $z$ and as
a result its solution is expressed via elliptic functions rather than simple
trigonometric or hyperbolic. 

If temperature is less then $T_c$, the parameter $\alpha
=\alpha _0(T-T_c)<0$ so that $\gamma /\alpha <0$ and solution of Eq.(\ref{22}) in
dimensionless variables (\ref{7}) subject to boundary conditions (\ref{5}) has the form
\begin{mathletters}
\begin{eqnarray}
\chi (z) &=&\frac 1\alpha \left\{ 1+\frac 1{\Delta _1}\left[ \left( \cos
\left( l_1\right) -\delta _{02}\sin \left( l_1\right) -1\right) \sin \left(
z_1\right) -\right. \right.  \nonumber \\
&&\ \ \left. \left. \left( \delta _{02}\cos \left( l_1\right) +\sin \left(
l_1\right) +\delta _{01}\right) \cos \left( z_1\right) \right] \right\} . 
\label{23a}  
\end{eqnarray}
Here

\begin{equation}
\Delta _1=(\delta _{01}+\delta _{02})\cos \left( l_1\right) +(1-\delta
_{01}\delta _{02})\sin \left( l_1\right)  \label{23}  
\end{equation}
\end{mathletters}
The expressions (\ref{23a}, \ref{23}) define the dielectric susceptibility profile in
paraelectric phase. Mean value of the susceptibility is found on the base of
Eq.(\ref{6}) and the integration yields:

\end{multicols}
\widetext
\noindent\rule{20.5pc}{0.1mm}\rule{0.1mm}{1.5mm}\hfill 
\begin{equation}
\overline{\chi }=-\frac 1\alpha \left[ \frac{2\left( 1-\cos \left(
l_1\right) \right) +(\delta _{01}+\delta _{02})\sin \left( l_1\right) }{%
l_1\left( (\delta _{01}+\delta _{02})\cos \left( l_1\right) +(1-\delta
_{01}\delta _{02})\sin \left( l_1\right) \right) }-1\right].  \label{24}
\end{equation}
\hfill\rule[-1.5mm]{0.1mm}{1.5mm}\rule{20.5pc}{0.1mm}
\begin{multicols}{2}
\narrowtext

Thickness dependence described by this expression is reported in Fig.8 by
left branches ($l<l_c$) of solid lines for several values of
dimensionless extrapolation lengths $\delta _{01}$ and $\delta _{02}$. It is
seen that $\chi (l\to l_c)\to \infty $ while approaching $l_c$
both from paraelectric and ferroelectric phase sides, however the
rate of this approaching is different and the value of susceptibility in
paraelectric phase is larger than that in ferroelectric phase. Temperature
dependence of mean dielectric susceptibility in paraelectric phase can be
calculated on the base of Eq.(\ref{24}) allowing for $l_1=\lambda \sqrt{%
1-\tau }$. The results of calculations are reported in Fig.9 by right 
branches ($T_c>T>T_{cl}$) of solid lines, the divergency at $T=T_{cl}$ as
well as linear dependence of $\chi _0/\overline{\chi }$ on $T/T_c$ being the
characteristic features.

When temperature is larger than $T_c$ (so that $\alpha >0$, $\gamma /\alpha
>0)$), $l_0$ becomes imaginary so that $z_1$ and $\delta
_{0i}$ are also imaginary. In this case the trigonometric functions
($\sin (\ldots )$ and $\cos (\ldots )$ )
in the solution (Eq.(\ref{23a})) of differential equation (\ref{22}) convert into
hyperbolic functions $i\sinh (\ldots )$ and $\cosh (\ldots )$.
In dimensionless variables (\ref{7}) the substitution $-\alpha %
 \to \alpha $ in Eqs.(\ref{23},\ref{24}) gives  the expressions for
distribution of susceptibility $\chi $ in paraelectic phase $T>T_c$: 
\end{multicols}
\widetext
\noindent\rule{20.5pc}{0.1mm}\rule{0.1mm}{1.5mm}\hfill

\begin{eqnarray}
\chi (z) &=&\frac 1\alpha \left( 1+\frac 1{\Delta _2}\left\{ \left[ \cosh
\left( l_1\right) +\delta _{02}\sinh \left( l_1\right) -1\right] \sinh
\left( z_1\right) -
\left[ \delta _{01}+\delta _{02}\cosh \left( l_1\right)
+\sinh \left( l_1\right) \right] \cosh \left( z_1\right) \right\} \right),
\label{25}
\end{eqnarray}
where 
\[
\Delta _2=(\delta _{01}+\delta _{02})\cosh \left( l_1\right) +(1+\delta
_{01}\delta _{02})\sinh \left( l_1\right) , 
\]
and mean value of susceptibility $\bar{\chi}$: 
\begin{equation}
\overline{\chi }=\frac 1\alpha \left[ 1-\frac{2\left( \cosh \left(
l_1\right) -1\right) +(\delta _{01}+\delta _{02})\sinh \left( l_1\right) }{%
(\delta _{01}+\delta _{02}^{^{\prime }})\cosh \left( l_1\right) +(1+\delta
_{01}\delta _{02})\sinh \left( l_1\right) }\right] ,T>T_c.  \label{26}
\end{equation}
\hfill\rule[-1.5mm]{0.1mm}{1.5mm}\rule{20.5pc}{0.1mm} 
\begin{multicols}{2}
\narrowtext

The thickness dependence of mean susceptibility (Eq.(\ref{26})) is reported in
Fig.10. It is seen from comparison of the Fig.10 and left branches of
the solid curves in Fig.8 that $\overline{\chi }$ behaviour in paraelectric
phase at $T>T_c$ and at $T_{cl}<T<T_c$ is strongly different: there is
neither divergency nor maximum at critical thickness, but $\overline{\chi }$ increases
slowly with thickness increase. As a matter of fact this difference is
related to the $l_c$ temperature dependence: $l_c=0$ at $T>T_c$. It is worth
stressing that the considered case $T>T_c$ corresponds to incipient
ferroelectric films, which are known to be in paraelectric phase with $T_c=0$%
, so that $T_{cl}=0$ too ($T_{cl}\leq T_c$ by definition).

\begin{figure}[th]
\vspace*{-3mm}
\centerline{\centerline{\psfig{figure=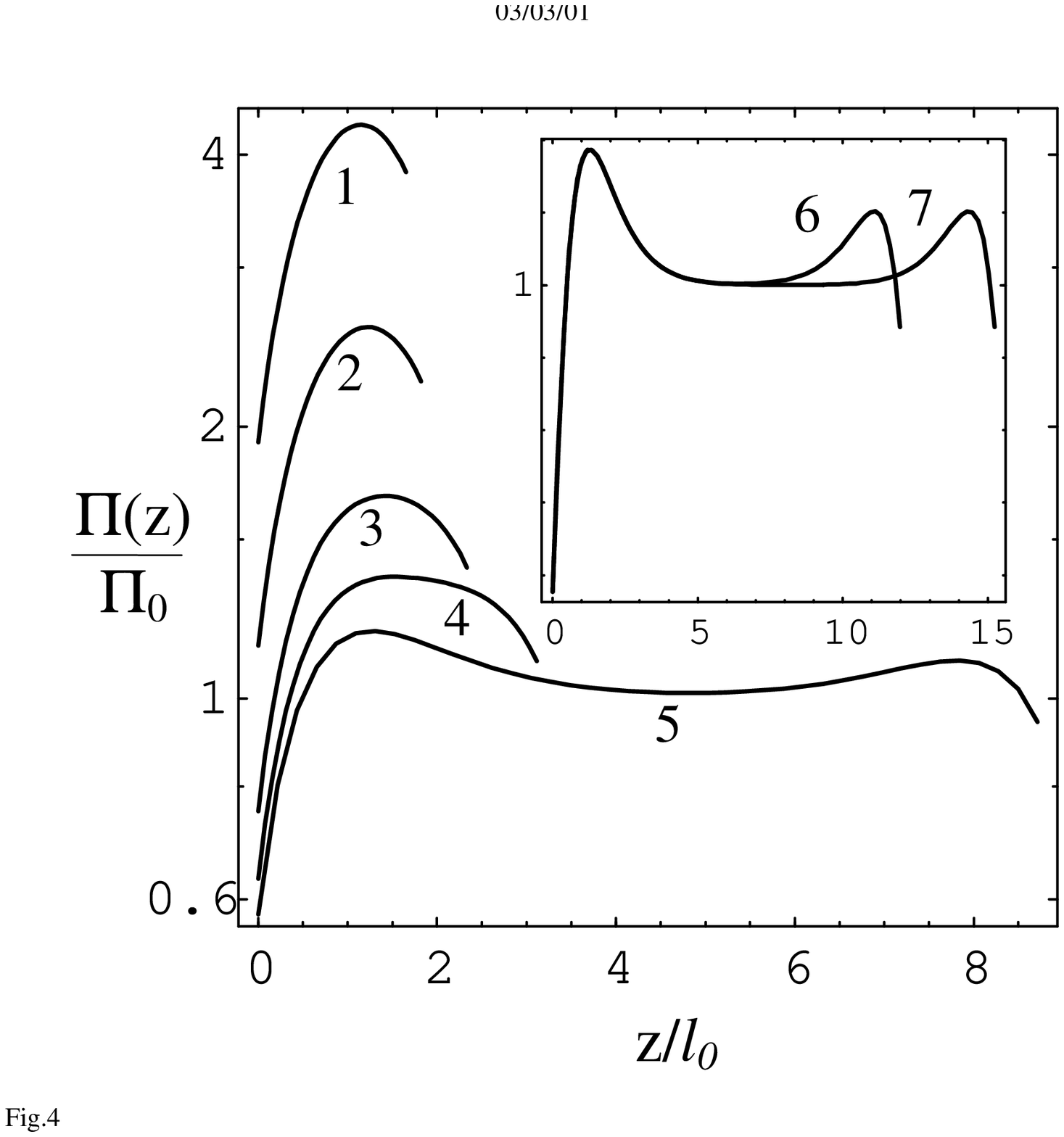,width=0.8\columnwidth}}}
%\vspace*{mm}
\caption{Pyroelectric coefficient profiles at $\protect\delta _{01}=0.5$, $%
\protect\delta _{02}=2$ for the following $m$ values: 0.04; 0.12; 0.33;
0.57; 0.99; $1-10^{-3}$; $1-10^{-4}$ respectively for the curves 1; 2; 3; 4;
5; 6; 7.}
\end{figure}

\begin{figure}[th]
\vspace*{-7mm}
\centerline{\centerline{\psfig{figure=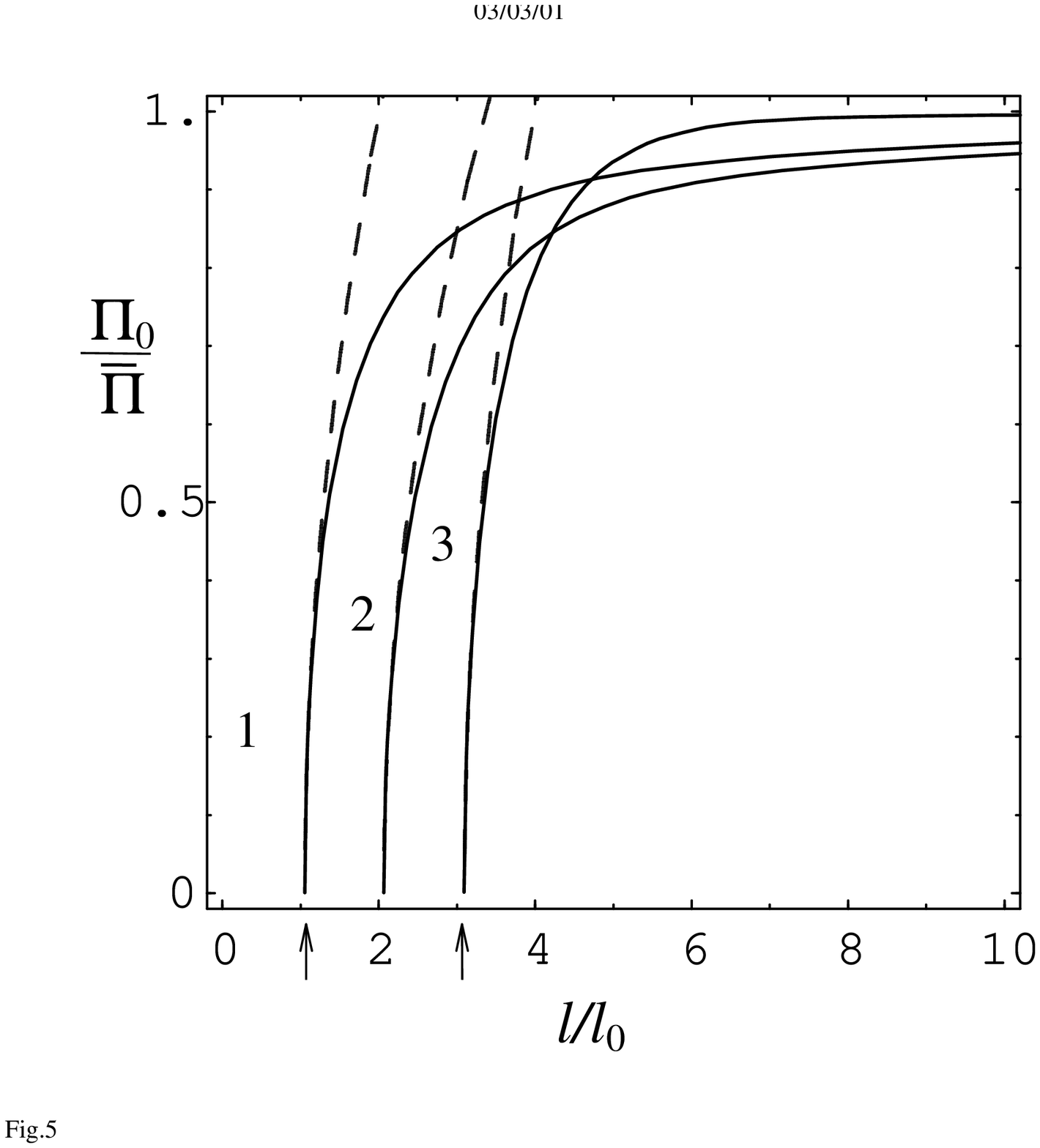,width=0.8\columnwidth}}}
%\vspace*{mm}
\caption{The thickness dependence of inverse mean pyroelectric coefficient
at $T<T_{cl}$ for the following $(\protect\delta _{01},\,\protect\delta
_{02})$ values: (1.5, 2), (0.6, 0.6) and (0.03, 0.02) for solid curves 1, 2
and 3 respectively. The dashed lines illustrate the description of inverse
mean pyroelectric coefficient by the approximate expression {(\ref{27})%
} with the same parameters as above. Arrows show the values of critical
thickness.}
\end{figure}

\section{Critical behaviour of pyroelectric coefficient and dielectric
susceptibility}
In this section we consider the behaviour of pyroelectric coefficient
and dielectric susceptibility in the vicinity of the thickness induced
ferroelectric phase transition in a film. This means that we have to
calculate the behaviour of these quantities at infinitesimal electric polarization,
i.e. parameter $m\to 0$. On the other hand $m\to 0$
corresponds to $l\to l_c$ or $T\to T_{cl}$.

The calculations can be performed by expansion of expressions (\ref{16}) and (\ref{21a})
for $\bar{\Pi}$ and $\bar{\chi}$ respectively into power series in $m$ up
to the first nonvanishing terms. Because of complex form of $\bar{\Pi} (m)$ and $\bar
{\chi }(m)$ (see Appendices 1, 2 and expressions (\ref{17a}, 
\ref{17b}, \ref{17c} and \ref{21b})) 
we put the detailed calculations in Appendix 3. The critical behaviour
of dielectric susceptibility while approaching the transition from
paraelectric phase is obtained directly from expression (\ref{24}) that is valid
for $T_{cl}\leq T<T_c$. These calculations lead to the following results:

Mean pyroelectric coefficient: 
\begin{mathletters}
\begin{equation}
\left. \Pi \right| _{l\to l_c+0}=\frac{C_{py}^l}{\sqrt{l_1-l_{1c}}}%
, l>l_c,  \label{27}
\end{equation}
\begin{equation}
\left. \Pi \right| _{\tau \to \tau _{cl}-0}=\frac{C_{py}^T}{\sqrt{%
\tau _{cl}-\tau }}, T<T_{cl};  \label{27b}  
\end{equation}
\end{mathletters}

\begin{figure}[th]
\vspace*{-5mm}
\centerline{\centerline{\psfig{figure=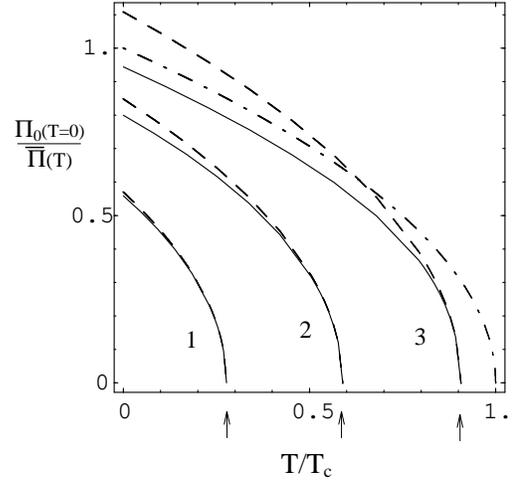,width=0.8\columnwidth}}}
%\vspace*{mm}
\caption{The temperature dependence of the inverse mean pyroelectric
coefficient for the films with following dimentionless thickness $\protect%
\lambda >\protect\lambda _c$: 2, 3 and 8 (see respectively solid lines 1, 2
and 3) at $d_1=0.5$, $d_2=2$. The dashed lines illustrate the description of
the inverse mean pyroelectric coefficient given by the approximate
expression {(\ref{27b})} with the same parameters as above. The
dash-dotted line illustrate the dependence of the inverse pyroelectric
coefficient of thick film $\Pi _0(T)$. Arrows show the values of critical
temperature.}
\end{figure}

Thickness dependence of mean dielectric susceptibility in paraelectric
phase: 
\begin{mathletters}
\begin{equation}
\left. \overline{\chi }\right| _{l\to l_c-0}=\frac{2C_{ch}^l}{%
l_{1c}-l_1}, T<T_c, l<l_c,  \label{28a} 
\end{equation}
and in ferroelectric phase: 
\begin{equation}
\left. \overline{\chi }\right| _{l\to l_c+0}=\frac{C_{ch}^l}{%
l_1-l_{1c}}, T<T_{cl}, l>l_c;  \label{28}  
\end{equation}
\end{mathletters}
Temperature dependence of mean dielectric susceptibility in ferroelectric
phase: 
\begin{mathletters}
\begin{equation}
\left. \bar{\chi}\right| _{\tau \to \tau _{cl}-0}=\frac{C_{ch}^T}{%
\tau _c-\tau }, l>l_c, T<T_{cl},  \label{29a}
\end{equation}
and in paraelectric phase: 
\begin{equation}
\left. \bar{\chi}\right| _{\tau \to \tau _c+0}=\frac{2C_{ch}^T}{\tau
-\tau _c}, l>l_c, T>T_{cl},  \label{29}
\end{equation}
\end{mathletters}
where constants $C_{py}^l,\,C_{py}^T,\,C_{ch}^l,\,C_{ch}^T$ can be found in
Appendix 3.

\begin{figure}[th]
\vspace*{-3mm}
\centerline{\centerline{\psfig{figure=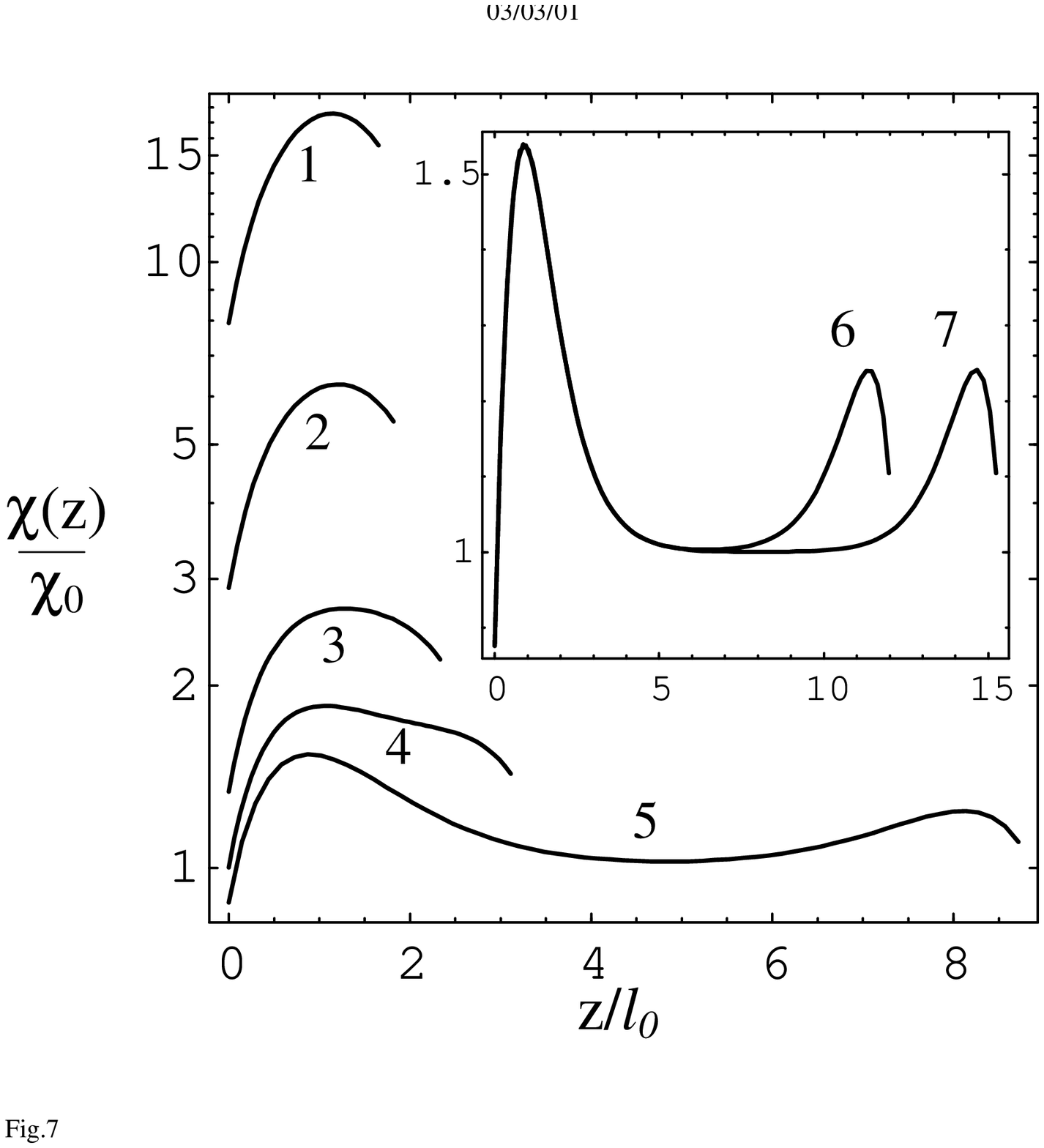,width=0.8\columnwidth}}}
%\vspace*{mm}
\caption{Dielectric susceptibility profiles in ferroelectric phase for the
same parameters as those in Fig.4.}
\end{figure}

\begin{figure}[th]
\vspace*{-5mm}
\centerline{\centerline{\psfig{figure=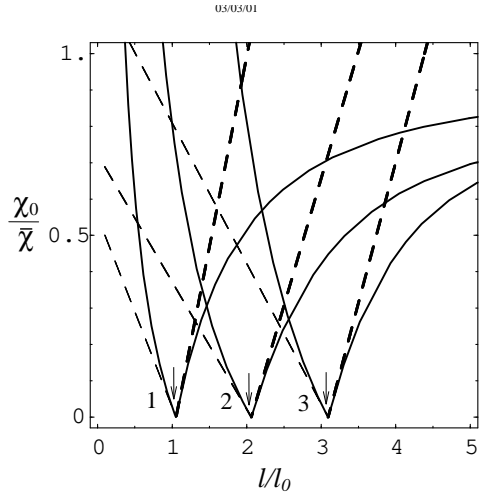,width=0.8\columnwidth}}}
%\vspace*{mm}
\caption{The thickness dependence of inverse mean dielectric susceptibility
at $T<T_c$ for the same parameters as those in Fig.5 (solid lines 1, 2 and
3). The dashed lines illustrate the description of inverse mean dielectric
susceptibility given by the approximate expressions {(\ref{28a},\ref{28})}.
Arrows show the values of critical thickness.}
\end{figure}

The dependencies given by expressions (\ref{27}-\ref{29}) are shown by dashed lines in
Figs.5, 6, 8 and by crosses in Fig.9. It is seen that for thin enough films
the expressions (\ref{27b}, \ref{29a}, \ref {29}) for the temperature critical behaviour,
(i.e. in the vicinity of $\tau \approx \tau _{cl}$) describe pretty good
the temperature dependence of $\overline{\chi }$ and $\overline{\Pi }$ far
enough from these points, while for the thickness critical behaviour the
approximate description (\ref{27}, \ref{28a}, \ref{28}) is good only 
in the close vicinity of $l_c$ (see
Figs. 5, 8). Note that the lines, corresponding to exact and approximate
solutions for curves 1 in Figs. 9
and 6 are almost the same, i.e. approximate formulas
perfectly fitted $\overline{\Pi }$ and $\overline{\chi }$ for the most thin
films. This is because thickness induced ferroelectric phase transition is
typical feature of thin films.

\begin{figure}[th]
\vspace*{-3mm}
\centerline{\centerline{\psfig{figure=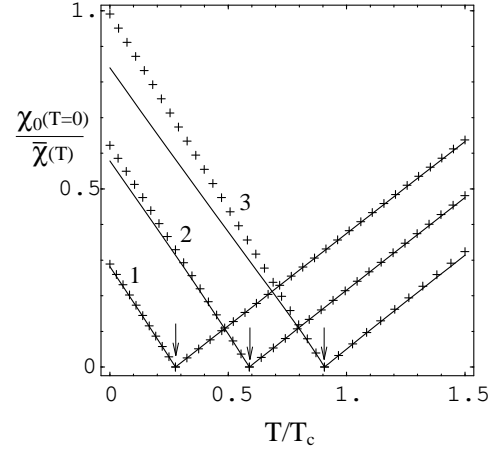,width=0.8\columnwidth}}}
%\vspace*{mm}
\caption{The temperature dependence of the inverse mean dielectric
susceptibility for the films with for the same parameters as those in Fig.6
(solid lines 1, 2 and 3). The crosses illustrate the dependence of inverse
mean dielectric susceptibility given by the approximate expressions (29a,
b). Arrows show the values of critical temperature.}
\end{figure}

It is important to emphasize that critical dependence of pyroelectric
coefficient (see Eq.(\ref{27b})) and dielectric susceptibility (Eq.(\ref{29a},\ref{29})) on
temperature is similar to that of bulk ferroelectrics.
% same as the ratio of
%constants in temperature dependence of mean susceptibility in ferroelectric
%(see Eq.(\ref{29a})) and paraelectric (see Eq.(\ref{29})) phases being equal 2.

The critical behaviour of the films with different thickness is related to
size effect of transition temperature $T_{cl}$. This effect is described by
expression (\ref{13a}) and is reported on Fig.3. Formula (\ref{13a}) can be simplified
for thick enough film and for thin films with thickness about critical one, namely 
\begin{mathletters}
\begin{eqnarray}
\tau _{cl}=1-\left( \frac \pi {\lambda +d_1+d_2}\right) ^2, \lambda \gg
1   \label{30a} \\
\tau _{cl}=1-\left( \frac{\lambda _c+d_1/\sqrt{1+d_1^2}+d_2/\sqrt{1+d_2^2}}{%
\lambda +d_1/\sqrt{1+d_1^2}+d_2/\sqrt{1+d_2^2}}\right) ^2, \nonumber \\
\lambda -\lambda _c\ll 1  \label{30}  
\end{eqnarray}
\end{mathletters}

It is seen from Eqs.(\ref{30}) that at $\lambda \to \infty ,$ i.e. $%
l\to \infty ,$ $\tau _{cl}\to 1$, i.e. $T_{cl}\to
T_c $, and at $\lambda \to \lambda _c$ $\tau _{cl}\to 0$ in
agreement with behaviour shown in Fig.3. It follows also from Fig.3 that
thick film limit can be obtained for finite $\lambda $ values even for small
enough $d_{1,2}$ (see curve 3, where $\tau _{cl}\approx 1$ at $%
\lambda \approx 15$ and $d_1=0,2$, $d_2=0,1$). To show how the approximate
expressions (\ref{30}) fit the exact curve (\ref{13a})
we show them in Fig.3 (for the parameters of curve 2 from that figure)
%the $\tau _{cl}=T_{cl}/T_c$ values calculated on the base of Eqs. (\ref{30a})
%and (\ref{30}) 
by dotted and dashed lines respectively. One can see the satisfactory agreement
between Eqs.(\ref{30}) and the curve 2, the fitting by expression (\ref{30}) is better
(for all values of $\lambda $) than that by Eq.(\ref{30a}).

\section{Discussion. Comparison with experiment}

We performed the analytical calculations of pyroelectric coefficient and
dielectric response profiles, their thickness and temperature dependence on
the base of solution of differential Lame equation. The considered boundary
conditions include different values of extrapolation lengths on the film
boundaries. To our mind this is more adequate for single film (not multilayer structire)
than the supposition used in the 
previous works, where the same extrapolation length (i.e. the same conditions on
both boundaries) have been used. In
addition, the majority of previous works (see e.g. \cite{8,9,10}) have been
devoted to numerical calculations of the properties of thin films made from
BaTiO$_3$ and PbTiO$_3$ materials. The analytical expressions
for description of size effects of pyroelectric coefficient, dielectric
response and polarization in ferroelectric thin films have the advantage of
their applicability to any film one can be interested in.

\begin{figure}[th]
\vspace*{-3mm}
\centerline{\centerline{\psfig{figure=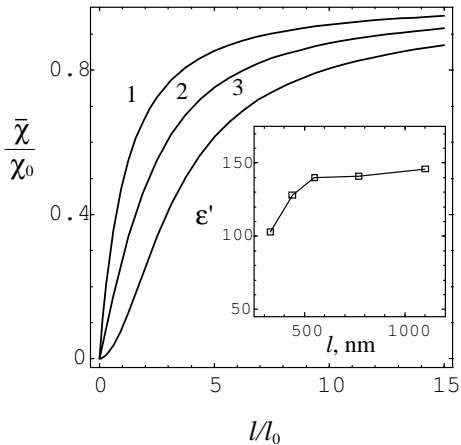,width=0.8\columnwidth}}}
%\vspace*{mm}
\caption{The calculated thickness dependence of mean dielectric
susceptibility at $T>T_c$ for the films with the same
parameters as those in Fig.8. The inset shows the observed 
thickness dependence of real part of permittivity (measured at 100 kHz) 
of $SrTiO_3$ film  \protect\cite{16}.}
\end{figure}

The theory predicts the divergency of dielectric susceptibility and
pyrocoefficient in the vicinity of thickness induced ferroelectric phase
transition. Temperature dependence of these quantities can be described by
Curie-Weiss law in wide enough temperature range while in thickness domain the
same dependencies $\chi \sim 1/\left| l-l_c\right| $ and $\Pi \sim 1/%
\sqrt{l-l_c}$ are valid only in the vicinity of critical thickness $l_c$.
 %the coefficients being dependent of film thickness
%(temperature domain) or on the temperature (thickness domain) (see Section 6).
To our mind these simple laws can be in real help for those 
interested in these properties measurements.

Unfortunately the results of measurements of size effects in ferroelectric
thin films are strongly restricted. This may be related to the
difficulties of thin films fabrication (especially with the thickness in
the vicinity of $l_c$ that can be several tens nanometers), to the influence of
electrodes on the film properties etc. (see e.g. \cite{16}). 

\begin{figure}[th]
\vspace*{-3mm}
\centerline{\centerline{\psfig{figure=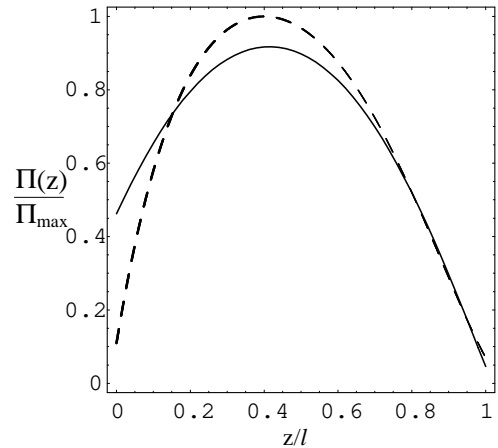,width=0.8\columnwidth}}}
%\vspace*{mm}
\caption{Pyrocoefficient profile calculated on the base of formulas (\ref{14}, \ref{15})
in section 5 at $m=0.1$; $\protect\delta _{01}=0.6$, $\protect\delta %
_{02}=0.05$ (solid line) and measured in PZT film \protect\cite{18} (dashed
line).}
\end{figure}

To overcome the latter difficulties the application of spectroscopic methods and 
optical and radiospectroscopy methods in particular seems to be desirable. Special methods
are needed also for investigations of the film properties profiles. The most
informative is ellipsometry method (see \cite{17} and ref. therein), 
making possible to measure optic refraction index profile. The
calculations of polarization profiles similar to those in Section 3 made it
possible to show that measured refraction index profile is related to the
film polarization profile \cite{17}. The special laser-intensity-modulation 
method (LIMM) can be applied to pyrocoefficient profile
measurements (see \cite{11,18}). Pyrocurrent coefficient profile extracted in 
\cite{18} from the pyrocurrent measurement by LIMM in 1 $\mu $m thick PbZr$%
_{0.75}$Ti$_{0.25}$O$_3$ film is shown by dashed line in Fig.11. It is
seen that it is looks really like some of the theoretical curves in Fig.4.
Detailed comparison of the theory with experiment is shown in Fig.11 where
solid line is theoretical curve with parameters $m=0.1$, $d_1=0.6$, $%
d_2=0.05$. We choose to fit the right "tail" of theoretical curve 
because of the absence of data related to high frequency
measurements so that it can be ambiguities in the experimental points
near left "tail" boundaries. However the procedure of pyrocoefficient
profile extraction from the measured (in broad
frequency range) pyrocurrent implies some numerical methods, the search for the
most effective one is still under way (see e.g. \cite{18}%
). For solution of this problem (that is the essential part of LIMM 
application to thin films) the calculations of $\Pi (z)$
profile seem to be important. On the other hand, spatial charge is present at
the near-electrode layer in real ferroelectic films, that leads to the
suppression of the ferroelectic long range order in this region \cite{BrLv}.
This phenomenon has not been taken into account in the present work.
Therefore both experimental and theoretical approaches need further
improvement for the better comparison of the theory with the experiment.

Unfortunately there is no measurements of pyroelectric coefficient
thickness dependence. Static dielectric susceptibility thickness dependence
measured recently in SrTiO$_3$ thin films \cite{16}; it is shown on the inset to
Fig.10. Since SrTiO$_3$ is an incipient ferroelectric, i.e. it is in
paraelectric phase for all temperatures ($T\geq 0$), $\bar \chi $
behaviour is described by Eq. (\ref{26}) (compare curves in the Fig.10 with
those in the inset) pretty good.

The theory prediction about the divergency of pyroelectric coefficient and static
dielectric susceptibility in the vicinity of thickness induced ferroelectric
phase transition in thin films is waiting for the experimental confirmation.

\appendix

\section{}
Let us rewrite the equations (\ref{4}) and (\ref{5a}) for
dielectric susceptibility $\chi (x)$ and pyroelectric coefficient $\Pi (x)$
with respect to expression (\ref{8}) for polarization distribution.
Introducing new variable $x=(z_1+z_0)/\sqrt{1+m}$ and denoting 
$\chi $ and $\Pi $ as $y$, one can find following equation for $y(x)$: 
\begin{eqnarray}
\frac{d^2y(x)}{dx^2}+\left( 1+m-6m\,y_0^2(x)\right) y(x) &=&g_y(x), 
\label{A1-1} \\
y_0(x) &=&{\rm sn}(x,m)  \nonumber
\end{eqnarray}
with the boundary conditions
\begin{eqnarray}
\left. \frac{dy(x)}{dx}\right| _{x=x_1}=\frac{\sqrt{1+m}}{\delta _{01}}%
\left. y(x)\right| _{_{x=x_1}}, \nonumber \\
\left. \frac{dy(x)}{dx}\right|
_{x=x_2}=-\frac{\sqrt{1+m}}{\delta _{02}}\left. y(x)\right| _{x=x_2} 
\label{A1-2}
\end{eqnarray}
General solution of the equation (\ref{A1-1}) (particular case of
Lame equation), is the following \cite{Erd}: 
\begin{mathletters}
\begin{eqnarray}
y(x) &=&C_1\,y_1(x)+C_2\,y_2(x)+y_3(x)  \label{A1-3} \\
y_1(x) &=&{\rm cn}(x,m)\,{\rm dn}(x,m)  \label{A1-3b} \\
y_2(x) &=&y_1(x)\int_0^x\frac{d\tilde x}{\left( y_1(\tilde x)\right) ^2} 
\label{A1-3c} \\
y_2(x) &=&y_1(x)\int_0^x\int_0^{\tilde x}\frac{g_y\left( \check x\right)
\,y_1\left( \check x\right) }{\left( y_1(\tilde x)\right) ^2}d\check
xd\tilde x  \label{A1-3d}
\end{eqnarray}
\end{mathletters}
The integral in the expression (\ref{A1-3c}) can be easily reduced into a standard
form (see \cite{G-R}), its further substitution to the formula (\ref{A1-3c})
gives the explicit form of the second fundamental solution of the equation
(\ref{A1-1}): 

\begin{eqnarray}
&&y_2(x)=\frac 1{1-m}\Biggl( \left( x-\frac{1+m}{1-m}E(am(x),m)\right)
y_1(x)+\nonumber \\
&&+y_0(x)\frac{1+m^2-m(1+m)\,y_0^2(x)}{1-m}\Biggr)  \label{A1-4}
\end{eqnarray}

To calculate the integral in the expression (\ref{A1-3d}) it is
necessary to know the explicit form of the function $g_y\left( x\right) $. This
function has the form (see Eqs.(\ref{4}, \ref{5a})): 
\begin{mathletters}
\begin{equation}
g_{pi}(x)=-2(1+m)\Pi _0\sqrt{\frac{2m}{1+m}}y_0(x)  \label{A1-5a}
\end{equation}
for the calculation of pyroelectric coefficient $\Pi (x)$ and 
\begin{equation}
g_{ch}(x)=-2(1+m)\chi _0  \label{A1-5b}
\end{equation}
\end{mathletters}
for the calculation of dielectric susceptibility $\chi (x)$. After
substitution (\ref{A1-5a},\ref{A1-5b}) to (\ref{A1-3d}) and integration over $\check x$ and $%
\tilde x$ \cite{G-R}, one can get the following expressions for the
inhomogeneous solution of equation (\ref{A1-1}): 

\begin{mathletters}
\begin{eqnarray}
&&y_3^{py}(x) =-\Pi _m\Biggl( \left( x-\frac{2\,E({\rm am}(x),m)}{1-m}\right)
y_1(x)+\nonumber \\
&&+y_0(x)\frac{1+m-2m\,\,y_0^2(x)}{1-m}\Biggr),  \nonumber \\
&&\Pi _m =\Pi _0\frac{\sqrt{2m\left( 1+m\right) }}{1-m}  \label{A1-6a}
\end{eqnarray}
and 
\begin{equation}
y_3^{ch}(x)=-2\left( 1+m\right) \chi _0\frac{1+m-2m\,\,y_0^2(x)}{\left(
1-m\right) ^2}  \label{A1-6b}
\end{equation}
\end{mathletters}

Constants $C_1,C_2$ are determined from the boundary conditions (\ref{A1-2}).
Substituting expression (\ref{A1-3}) into (\ref{A1-2}), one can obtain the system of
linear equations for calculation of $C_1$ and $C_2$. The solution of this
system gives following form for constants $C_1$ and $C_2$

\begin{mathletters}
\begin{eqnarray}
C_1 =\frac{B_1A_{22}-B_2A_{21}}\Delta, \nonumber \\   
C_2 =\frac{B_2A_{11}-B_1A_{12}}\Delta, \nonumber \\   
\Delta =A_{22}A_{11}-A_{21}A_{12}  \label{A1-7a} 
\end{eqnarray}
where 

\begin{eqnarray}
&&A_{11} =\frac{\sqrt{1+m}}{\delta _{01}}y_1(x_1)-y_1^{\prime }(x_1),\nonumber \\
&&A_{21}=\frac{\sqrt{1+m}}{\delta _{01}}y_2(x_1)-y_2^{\prime }(x_1),  \nonumber \\
&&A_{12}=\frac{\sqrt{1+m}}{\delta _{01}}y_1(x_2)+y_1^{\prime }(x_2),\nonumber \\
&&A_{22}=\frac{\sqrt{1+m}}{\delta _{01}}y_2(x_2)+y_2^{\prime }(x_2),  \nonumber \\
&&B_1 =y_3^{\prime }(x_1)-\frac{\sqrt{1+m}}{\delta _{01}}y_3(x_1),\nonumber \\
&&B_2=-y_3^{\prime }(x_2)-\frac{\sqrt{1+m}}{\delta _{01}}y_3(x_2)  
 \label{A1-7b}
\end{eqnarray}
\end{mathletters}

Using the explicit form of solutions $y_1(x)$, $y_2(x)$, $y_3(x)$ (\ref{A1-3b}, \ref{A1-4},
\ref{A1-6a}, \ref{A1-6b}), values of $x_1$, $x_2$ and notations for $e_1$ and $e_2$
\begin{eqnarray*}
x_1 &=&F(\arcsin (f_1),m), x_2=2K(m)-F(\arcsin (f_2),m) \\
e_1 &=&E(\arcsin (f_1),m), e_2=2E(m)-E(\arcsin (f_2),m) \\
f_i &=&\sqrt{\frac{1+m}{2m}\left( 1+\frac 1{\delta _{0i}^2}-\sqrt{\left(
1+\frac 1{\delta _{0i}^2}\right) ^2-\frac{4m}{\left( 1+m\right) ^2}}\right) }
\end{eqnarray*}
it is easy to obtain following expression for $A_{ij}$ and $B_i$ 

\begin{eqnarray}
&&A_{11} =f_1\left( (1+m)\left( 1+\frac 1{\delta _{01}^2}\right)
-2m\,f_1^2\right) , \nonumber \\ 
&&A_{12} =-f_2\left( (1+m)\left( 1+\frac 1{\delta _{02}^2}\right)
-2m\,f_2^2\right) ,  \nonumber \\
&&A_{2i} =\frac 1{1-m}\Biggl( f_i\frac{\sqrt{1+m}}{\delta _{0i}}\frac{%
2m+m(1+m)f_i^2}{1-m}+\nonumber \\
&&+A_{1i}\left( x_i-\frac{1+m}{1-m}e_i \right)\Biggr) , 
\label{A1-8} \\
&&B_i^{py} =\Pi _m\Biggl( f_i\frac{\sqrt{1+m}}{\delta _{0i}}\frac{%
1+m+2m\,f_i^2}{1-m}+\nonumber \\
&&+A_{1i}\left( x_i-\frac{2\,e_i}{1-m}\right) \Biggr) , 
\label{A1-9} \\
&&B_i^{ch} =\chi _0\frac{2\left( 1+m\right) \sqrt{1+m}}{\left( 1-m\right)
^2\delta _{0i}}\left( 1+m+2m\,f_i^2\right), \nonumber \\
&&i=1,2.  \label{A1-10} 
\end{eqnarray}

This permits to get final form of coefficients $C_1^{py}$, $C_2^{py}$ with the help
of expressions (\ref{A1-7a}, \ref{A1-8}, \ref{A1-9}) 
and coefficients $C_1^{py}$, $C_2^{py}$ with
the help of expressions (\ref{A1-7a}, \ref{A1-8}, \ref{A1-10}).

\section{}

Mean pyroelectric coefficient $\bar{\Pi}$ and dielectric susceptibility $%
\bar{\chi}$ are introduced in the expression (\ref{6}). After changing of variables
in the integrals (\ref{6}), the mean value of function $g(x)$ can be written as
\[
\bar{g}=\frac 1{x_2-x_1}\int\limits_{x_1}^{x_2}g(x)\,dx 
\]
Using expressions (\ref{14}, \ref{19}) for $\chi (x)$ and $\Pi (x)$, we 
can write their mean values as follows 
\begin{mathletters}
\begin{eqnarray}
\bar{\Pi} &=&C_1^{py}\bar{y}_1+C_2^{py}\bar{y}_2+\bar{y}_3^{py}  \label{A2-1a}
\\
\bar{\chi} &=&C_1^{ch}\bar{y}_1+C_2^{ch}\bar{y}_2+\bar{y}_3^{ch}  \label{A2-1b}
\end{eqnarray}
\end{mathletters}
To calculate $\bar{\Pi}$ and $\bar{\chi}$ it is necessary to
integrate all the solutions of equation (\ref{A1-1}) over $x$. Taking into account
the explicit form (\ref{A1-3b}, \ref{A1-4}, \ref{A1-6a}, \ref{A1-6b}) 
of $y_1(x)$, $y_2(x)$, $y_3(x)$, it is easy to reduce the integrals to the standard
form and to find that (see \cite{G-R})
\[
\int y_1(x)dx=y_0(x), 
\]

\begin{eqnarray*}
&&\int y_2(x)dx=\\
&&=\frac 1{1-m}\left( \left( x-\frac{1+m}{1-m}E({\rm am}(x),m)\right)
y_0(x)-\frac{1+m}{1-m}y_1(x)\right) , \\
&&\int y_3^{py}(x)dx=\\
&&=-\Pi _m\left( \left( x-\frac{2\,E({\rm am}(x),m)}{1-m}\right)
y_0(x)-\frac{2\,y_1(x)}{1-m}\right),\\ 
&&\int y_3^{ch}(x)dx=2\,\chi _0\frac{1+m}{1-m}\left( x-\frac{2\,E({\rm am}(x),m)}{1-m%
}\right) . 
\end{eqnarray*}

These expressions along with values of $x_1$ and $x_2$ permit to get the mean
values of $y_1(x)$, $y_2(x)$, $y_3(x)$ in the following form

\begin{mathletters}
\begin{eqnarray}
&&\overline{y}_{1} =\frac{f_{2}-f_{1}}{x_{2}-x_{1}},  \label{A2-2} \\
&&\overline{y}_{2} =\frac{1}{\left( x_{2}-x_{1}\right)\left( 1-m\right)}\times \nonumber \\
&&\times \Biggl[f_{2}\left( x_{2}-\frac{1+m}{1-m}\left( e_{2}-\frac{\sqrt{1+m}}{\delta _{02}%
}\right) \right) -  \nonumber \\
&&-f_{1}x_{1}-\frac{1+m}{1-m}\left( e_{1}+\frac{\sqrt{1+m}}{\delta _{01}}%
\right) \Biggr],  \label{A2-3}\\
&&\overline{y}_{3}^{py} =\frac{\Pi _{m}}{\left( x_{2}-x_{1}\right) \left( 1-m\right)}
\times \nonumber \\
&&\times \Biggl[f_{2}\left( 2\left( e_{2}-\frac{\sqrt{1+m}}{\delta _{02}}\right) 
-\left( 1-m\right) x_{2}\right) -\nonumber \\
&&-f_{1}\left(
2\left( e_{1}+\frac{\sqrt{1+m}}{\delta _{01}}\right) -\left( 1-m\right)
x_{1}\right)\Biggr], \label{A2-4a} \\
&&\overline{y}_{3}^{ch} =\frac{\chi _{0}}{x_{2}-x_{1}}\frac{1+m}{1-m}\left(
x_{2}-x_{1}-\frac{2\left( e_{2}-e_{1}\right) }{1-m}\right) .  \label{A2-4b}
\end{eqnarray}

\subsection{Ferroelectric phase}

For the investigation of behavior of mean pyroelectric coefficient 
$\bar {\Pi } $ and dielectric susceptibility $\bar \chi $ in the vicinity of thickness
induced phase transition, we are to expand $\bar \Pi (m)$ and $\bar \chi (m)$
at $m\to 0$. Using the series for elliptic integrals \cite{G-R},
one can expand $A_{ij}$, $B_i$ and $\bar y_i$ (see (\ref{A1-8}, 
\ref{A1-9}, \ref{A1-10}) and (\ref{A2-2}, \ref{A1-3}, \ref{A2-4a}, \ref
{A2-4b})) to the first nonvanishing terms and obtain following relations
\end{mathletters}
\begin{mathletters}
\begin{eqnarray}
&&\left. \bar \chi \right| _{m\to 0} =\chi _0\frac{4\left( \frac{S_1}{\delta
_{01}}+\frac{S_2}{\delta _{02}}\right) ^2}{3\Delta _0l_{1c}}\frac 1m
\label{A3-1a} \\
&&\left. \bar \Pi \right| _{m\to 0} =\nonumber \\
&&=\Pi _0\frac{2\sqrt{2}\left( l_{1c}+%
\frac{S_1^2}{\delta _{01}}+\frac{S_2^2}{\delta _{02}}\right) \left( \frac{S_1%
}{\delta _{01}}+\frac{S_2}{\delta _{02}}\right) }{3\Delta _0l_{1c}}\frac 1{%
\sqrt{m}}  \label{A3-1b}
\end{eqnarray}
where 
\end{mathletters}
\begin{eqnarray*}
S_i=\frac{\delta _{0i}}{\sqrt{1+\delta _{0i}^2}}, i=1,2; \\
l_{1c}=\pi -\arcsin \left( S_1\right) -\arcsin \left( S_2\right)
\end{eqnarray*}
\[
\Delta _0=\left( l_{1c}+\frac{S_1^2}{\delta _{01}}\left( 1+\frac 23%
S_1^2\right) +\frac{S_2^2}{\delta _{02}}\left( 1+\frac 23S_2^2\right)
\right) 
\]
The dependence of dimensionless thickness $l_1 (m)$ has the form 
\begin{equation}
l_1=\sqrt{1+m}\left( x_2-x_1\right) .  \label{A3-2}
\end{equation}
The expansion over $m$ gives following relation 
\[
\left. l_1\right| _{m\to 0}=l_{1c}+\frac 34\Delta _0\,m 
\]
Substitution of this expression into (\ref{A3-1a}, \ref{A3-1b}) gives the
structure of mean pyroelectric coefficient $\bar \Pi $ and dielectric
susceptibility $\bar \chi $ in the vicinity of thickness induced phase
transition $l_1\to l_{1c}+0:$ 
\begin{mathletters}
\begin{eqnarray}
&&\left. \bar \chi \right| _{l_1\to l_{1c}+0} =\frac{C_{ch}^l}{l_1-l_{1c}},
\label{A3-3a} \\
&&C_{ch}^l =\chi _0\frac{\left( \frac{S_1}{\delta _{01}}+\frac{S_2}{\delta
_{02}}\right) ^2}{l_{1c}}  \label{A3-3b}
\end{eqnarray}
\end{mathletters}
\begin{mathletters}
\begin{eqnarray}
&&\left. \bar \Pi \right| _{l_1\to l_{1c}+0} =\frac{C_{py}^l}{\sqrt{%
l_1-l_{1c}}},  \label{A3-4a} \\
&&C_{py}^l =\Pi _0\frac{\sqrt{2}\left( l_{1c}+\frac{S_1^2}{\delta _{01}}+%
\frac{S_2^2}{\delta _{02}}\right) \left( \frac{S_1}{\delta _{01}}+\frac{S_2}{%
\delta _{02}}\right) }{\sqrt{3\Delta _0}l_{1c}}  \label{A3-4b}
\end{eqnarray}
The temperature dependence of $\bar{\Pi}$ and $\bar{\chi}$ in the vicinity of
the phase transition can also be derived from (\ref{A3-2}) and (\ref{A3-1a}, 
\ref{A3-1b}) with respect to obvious form of $l_1$ and $\delta _{0i}$.
Introducing new notations 
\end{mathletters}
\begin{eqnarray}
l_1 &=&\lambda \sqrt{1-\tau }, \lambda =l_1(T=0),  \nonumber \\
\delta _{0i} &=&d_i\sqrt{1-\tau }, d_i=\delta _{0i}(T=0),  \label{A3-5} \\
\lambda _c &=&l_{1c}(T=0), \tau =\frac T{T_c},  \nonumber
\end{eqnarray}
and using equation for critical temperature $\tau _{cl}=T_{cl}/T_c$ 

\begin{eqnarray*}
&&\lambda \sqrt{1-\tau _{cl}}=
\pi -\arcsin \left( \frac{d_1\sqrt{1-\tau _{cl}}%
}{\sqrt{1+d_1^2\left( 1-\tau _{cl}\right) }}\right) -\\
&&-\arcsin \left( \frac{d_2%
\sqrt{1-\tau _{cl}}}{\sqrt{1+d_2^2\left( 1-\tau _{cl}\right) }}\right) 
\end{eqnarray*}
we can expand the expression (\ref{A3-2}) and get following series up to the
first order in $m$  
\[
\left. \tau \right| _{m\to 0}=\tau _c-\frac{\frac 32\left( 1-\tau
_{cl}\right) \Delta _0\left( \tau =\tau _{cl}\right) }{\lambda _c\sqrt{%
1-\tau _{cl}}+\frac{d_1\sqrt{1-\tau _{cl}}}{1+d_1^2\left( 1-\tau
_{cl}\right) }+\frac{d_2\sqrt{1-\tau _{cl}}}{1+d_2^2\left( 1-\tau
_{cl}\right) }}\,m. 
\]
Applying this expansion to (\ref{A3-1a}, \ref{A3-1b}), we can rewrite it as
follows 
\begin{mathletters}
\begin{eqnarray}
&&\left. \bar{\chi}\right| _{\tau \to \tau _{cl}-0} =\frac{C_{ch}^T}{\tau
_{cl}-\tau },  \label{A3-6a} \\
&&C_{ch}^T =\frac{2N^2}{\lambda _c\sqrt{1-\tau _{cl}}D}\chi _0(T=0),
\label{A3-6b}
\end{eqnarray}
\end{mathletters}
\begin{mathletters}
\begin{eqnarray}
&&\left. \bar{\Pi}\right| _{\tau \to \tau _c-0} =\frac{C_{py}^T}{\sqrt{\tau
_{cl}-\tau }},  \label{A3-7a} \\
&&C_{py}^T =\sqrt{\frac{4D}{3G}}\frac N{\lambda _c\sqrt{1-\tau _{cl}}}\Pi
_0(T=0),  \label{A3-7b}
\end{eqnarray}
where 
\end{mathletters}
\[
N=\frac 1{\sqrt{1+d_1^2\left( 1-\tau _{cl}\right) }}+\frac 1{\sqrt{%
1+d_2^2\left( 1-\tau _{cl}\right) }}, 
\]
\[
D=\lambda _c\sqrt{1-\tau _{cl}}+\frac{d_1\sqrt{1-\tau _{cl}}}{1+d_1^2\left(
1-\tau _{cl}\right) }+\frac{d_2\sqrt{1-\tau _{cl}}}{1+d_2^2\left( 1-\tau
_{cl}\right) }, 
\]
\[
G=D+\frac 23\frac{\left( d_1\sqrt{1-\tau _{cl}}\right) ^3}{\left(
1+d_1^2\left( 1-\tau _{cl}\right) \right) ^2}+\frac 23\frac{\left( d_2\sqrt{%
1-\tau _{cl}}\right) ^3}{\left( 1+d_2^2\left( 1-\tau _{cl}\right) \right) ^2}%
. 
\]

\subsection{Paraelectric phase}

Mean dielectric susceptibility $\bar{\chi}$ in paraelectric phase (i.e. at
temperature above $T_c$ or at $l_1<l_{1c}$) has the form 
\begin{equation}
\bar{\chi}=2\chi _0\left( \frac 1{l_1}\frac{\left( \delta _{01}+\delta
_{02}\right) \sin \left( l_1\right) +2\left( 1-\cos \left( l_1\right)
\right) }{\left( 1-\delta _{01}\delta _{02}\right) \sin \left( l_1\right)
+\left( \delta _{01}+\delta _{02}\right) \cos \left( l_1\right) }-1\right)
\label{A3-8}
\end{equation}
Expanding (\ref{A3-8}) in terms of $\left( l_1-l_{1c}\right) $ one can write
the asymptotic form of $\bar{\chi}$ at $l_1 \to l_{1c}$
\begin{equation}
\left. \bar{\chi}\right| _{l_1\to l_{1c}-0}=\frac{2C_{ch}^l}{l_{1c}-l_1},
\label{A3-9}
\end{equation}
where $C_{ch}^l$ is given by expression (\ref{A3-3b}). Using
the temperature dependence (\ref{A3-5}) of $l_1$ and $\delta _{0i}$, 
it is easy to rewrite (\ref{A3-8}) as follows 
\begin{equation}
\left. \bar{\chi}\right| _{\tau \to \tau _{cl}+0}=\frac{2C_{ch}^T}{\tau
-\tau _{cl}},  \label{A3-10}
\end{equation}
where coefficient $C_{ch}^T$ is given by the expression (\ref{A3-6b}).

\end{multicols}

\end{document}